\documentclass[10pt]{article}
\usepackage{geometry}                % See geometry.pdf to learn the layout options. There are lots.
\usepackage{graphicx}
\usepackage{subfig}
\usepackage{amssymb}
\usepackage{amsmath}
\usepackage{hyperref}
\usepackage{epstopdf}
\usepackage{caption}
\usepackage[round]{natbib}
\newcommand{\vecb}{\boldsymbol{\mathbf{b}}}

\newcommand{\vecx}{\boldsymbol{\mathbf{x}}}
\newcommand{\vecy}{\boldsymbol{\mathbf{y}}}
\newcommand{\vecz}{\boldsymbol{\mathbf{z}}}

\newcommand{\tv}{\boldsymbol{\theta}}
\newcommand{\mv}{\boldsymbol{\mu}}

\newcommand{\Sv}{\boldsymbol{\Sigma}}

\newcommand{\Pv}{\boldsymbol{\Phi}}

\newcommand{\bv}{\boldsymbol{\beta}}

\title{Information-adaptive clinical trials:\\a selective recruitment design}
\author{James E. Barrett\\\normalsize{University College London, London, U.K.}}
\date{March 28, 2016}
\begin{document}
\maketitle

\begin{abstract}
We propose a novel adaptive design for clinical trials with time-to-event outcomes and covariates (which may consist of or include biomarkers). Our method is based on the expected entropy of the posterior distribution of a proportional hazards model. The expected entropy is evaluated as a function of a patient's covariates, and the information gained due to a patient is defined as the decrease in the corresponding entropy. Candidate patients are only recruited onto the trial if they are likely to provide sufficient information. Patients with covariates that are deemed uninformative are filtered out. A special case is where all patients are recruited, and we determine the optimal treatment arm allocation. This adaptive design has the advantage of potentially elucidating the relationship between covariates, treatments, and survival probabilities using fewer patients, albeit at the cost of rejecting some candidates. We assess the performance of our adaptive design using data from the German Breast Cancer Study group and numerical simulations of a biomarker validation trial.
\end{abstract}

%ÇÇÇÇÇÇÇÇÇÇÇÇÇÇÇÇÇÇÇÇÇÇÇÇÇÇÇÇÇÇÇÇÇÇÇÇÇÇÇÇÇÇÇÇÇÇÇÇÇÇÇÇÇÇÇÇÇÇÇÇÇ%
%
%
\section{Introduction}
%
%
%ÇÇÇÇÇÇÇÇÇÇÇÇÇÇÇÇÇÇÇÇÇÇÇÇÇÇÇÇÇÇÇÇÇÇÇÇÇÇÇÇÇÇÇÇÇÇÇÇÇÇÇÇÇÇÇÇÇÇÇÇÇ%

Adaptive clinical trials offer a potentially more efficient and ethical way to conduct clinical trials. Covariate-adaptive designs try to ensure that the distributions of covariates across different arms are balanced, thus resulting in more comparable cohorts on each arm \citep{pocock1975sequential, taves1974minimization}. Response-adaptive randomisation attempts to allocate more patients to the effective treatment arms. As the trial progresses and more information is acquired on the efficacies of each treatment arm the allocation probabilities shift towards the more effective treatments. \citet{zhang2007response} develop an optimal response-adaptive design under exponential and Weibull parametric models for time-to-event outcomes. See \citet{yin2013clinical} for a good overview of adaptive designs.

We regard the primary goal of a clinical trial as establishing a statistical relationship between covariates, treatments, and survival outcomes. As we will show, not all patients on a trial provide the same amount of statistical information. Some covariate values are more informative than others. In addition, the informativeness of a covariate value will depend on what has been observed so far in the trial. As an example, consider two scenarios where a patient with particular covariate values is available for recruitment. In the first scenario another patient with precisely the same covariate values has already been recruited. In the second scenario suppose the candidate's covariates come from a region of covariate space that has not previously been sampled. Intuitively we expect the candidate to be more informative in the second scenario since they provide access to previously unobserved covariates values and outcomes.

Our aim in this paper is to address a practical question: given limited resources and the observation that not all patients are equally informative, what is the optimal way to conduct a clinical trial? We propose that it may be advantageous to selectively recruit and allocate patients on the basis of how much information they are likely to provide. Covariates are measured for candidate patients, and based on those values and what has been inferred from the trial up to that point a recruitment probability is computed. In other words, we  \emph{filter} out patients that are unlikely to significantly reduce the uncertainty surrounding model parameters. 

\emph{Predictive} biomarkers, which indicate whether a patient is likely to respond well to a particular treatment or not, are increasingly useful in the drive towards personalised medicine and targeted therapy. A potential application of our selective-recruitment design would be to validate a biomarker by looking at treatment-biomarker interaction terms in a proportional hazards model. We test this using numerical simulations. \citet{sargent2005clinical} discuss alternative adaptive designs for validating predictive biomarkers.

Our filtering approach is similar in spirit to some existing designs. \citet{freidlin2005adaptive} propose a trial design which attempts to find a gene signature that will identify a subset of `sensitive' patients who are more likely to respond to the treatment. In a randomised discontinuation design \citep{rosner2002randomized} patients who fail to respond to a treatment in the first phase of the trial are dropped from the second part, thereby isolating a responsive subset of patients with a stronger statistical signal. Another type of trial known as `enrichment designs' \citep{temple2010enrichment} enrich the recruited cohort with patients who are more likely to have the event of interest. For example, patients with a particular biomarker. Given that more events of interest are observed greater statistical power can be achieved within the enriched cohort.

We assume a proportional hazards model with a constant baseline hazard rate. The entropy of the posterior distribution is a useful way to quantify our uncertainty regarding the model parameters. As the trial progresses, and the space of plausible parameter values shrinks, the entropy decreases. The \emph{informativeness} of a candidate is defined as the reduction in expected entropy in the hypothetical scenario where they are added to the cohort of existing recruits. The \emph{ideal} candidate at time $t$ is defined as the patient that would achieve the greatest possible reduction in expected entropy. By comparing the current candidate to the ideal candidate we can obtain a recruitment probability. The posterior is constructed using outcomes from all patients accrued up until time $t$. Patients who have not experienced any events are considered to be right-censored. Therefore, the recruitment probability changes dynamically as more events and patients are observed. An arm allocation probability can also be computed based on which arm has the lowest expected entropy. We also implement this in a more traditional setting where all candidates are recruited.

In Section \ref{sec:protocol} we provide the mathematical details and describe some approximations which are required. Results from experimental data generated by the German Breast Cancer Study group and numerical simulations are presented in Sections \ref{sec:results:gbcs} and \ref{sec:results:num} respectively. Discussion on the practical applicability of our approach and concluding remarks are given in Section \ref{sec:disc}.

%ÇÇÇÇÇÇÇÇÇÇÇÇÇÇÇÇÇÇÇÇÇÇÇÇÇÇÇÇÇÇÇÇÇÇÇÇÇÇÇÇÇÇÇÇÇÇÇÇÇÇÇÇÇÇÇÇÇÇÇÇÇ%
%
%
\section{An information based adaptive protocol}
\label{sec:protocol}
%
%
%ÇÇÇÇÇÇÇÇÇÇÇÇÇÇÇÇÇÇÇÇÇÇÇÇÇÇÇÇÇÇÇÇÇÇÇÇÇÇÇÇÇÇÇÇÇÇÇÇÇÇÇÇÇÇÇÇÇÇÇÇÇ%

%°°°°°°°°°°°°°°°°°°°°°°°°°°°°°°°°°°°°°°%
\subsection{Proportional hazards model}
\label{sec:ph}
%°°°°°°°°°°°°°°°°°°°°°°°°°°°°°°°°°°°°°°%

Suppose that $N_t$ patients have been recruited onto the trial at time $t$. Observed data are denoted by $D_t = \{(\vecx_1,t_1,\Delta_1),\ldots,(\vecx_{N_t},t_{N_t},\Delta_{N_t})\}$ where $\vecx_i\in\mathbb{R}^d$ is a vector of covariates for patient $i$ (this vector may include biomarker values or treatment indicator variables). If patient $i$ is censored then $\Delta_i=0$ and $t_i$ is the time of censoring, otherwise the primary event occurred at time $t_i$ and $\Delta_i=1$. Patients who have not experienced any event by $t$ are considered right censored. We assume a proportional hazards model with a constant baseline hazard rate $\lambda\in(0,\infty)$:
\begin{equation}
h(t_i|\vecx_i,\lambda,\bv) = \lambda e^{\bv\cdot\vecx_i} \quad \text{for $i=1,\ldots,N_t$}
\end{equation}
where $\bv\in\mathbb{R}^d$ is a vector of regression coefficients. The covariates are assumed to be drawn from a known population distribution $p(\vecx)$. The data likelihood is
\begin{equation}
p(D_t|\lambda,\bv) = \prod_{i=1}^{N_t} \left(\lambda e^{\bv\cdot\vecx_i}\right)^{\Delta_i}\exp(-\lambda t_ie^{\bv\cdot\vecx_i})p(\vecx_i).
\end{equation}
Using Bayes' rule we can write the posterior as
\begin{equation}
p(\lambda,\bv|D_t,\tv) = \frac{p(D_t|\lambda,\bv)p(\lambda|\tv)p(\bv|\tv)}{p(D_t|\tv)}
\label{eq:posterior}
\end{equation}
where $p(D_t|\tv)$ is the marginal likelihood. The vector $\tv$ contains hyperparameters that are required for the prior distributions. For the prior over $\lambda$ we choose $\lambda \sim \text{Gamma}(\kappa_0,\chi_0)$, with shape and scale hyperparameters $\kappa_0$ and $\chi_0$ respectively, and $\bv \sim \mathcal{N}(0,\alpha_0^2I)$. The value of $\tv=(\kappa_0,\chi_0,\alpha_0^2)$ is fixed and we will henceforth drop the dependence on $\tv$ for the sake of notational compactness.

%°°°°°°°°°°°°°°°°°°°°°°°°°°°°°°°°°°°°°°%
\subsection{Entropy as a measure of patient informativeness}
%°°°°°°°°°°°°°°°°°°°°°°°°°°°°°°°°°°°°°°%

At time $t$ we have recruited $N_t$ patients onto the trial. Suppose that a candidate patient with covariates $\vecx^*$ has presented and we wish to estimate how much information we expect the candidate to provide if they are to be recruited. The information gain is defined as the reduction in the expected entropy of the posterior (\ref{eq:posterior}). The entropy is defined as
\begin{equation}
h(D_t) = -\left<\log p(\lambda,\bv|D_t)\right>_{p(\lambda,\bv|D_t)}.
\label{eq:entropy}
\end{equation}
The notation $\left<\cdots\right>_p$ denotes the expectation with respect to the density $p$. We then add the candidate to the existing cohort and take the expectation with respect to the unknown $t^*$:
\begin{equation}
H(\vecx^*|D_t) = \left< h(D_t\cup\{\vecx^*,t^*\})\right>_{p(t^*|\vecx^*,D_t)}
\label{eq:energy}
\end{equation}
where the argument of $h$ is the union of $D_t$ and the additional uncensored observation $\{\vecx^*,t^*\}$ and where
\begin{equation}
p(t^*|\vecx^*,D_t) = \left<p(t^*|\vecx^*,\lambda,\bv)\right>_{p(\lambda,\bv|D_t)}.
\label{eq:pred}
\end{equation}
The time-to-event density is $p(t^*|\vecx^*,\lambda,\bv) = \lambda e^{\bv\cdot\vecx^*} \text{exp}(-\lambda t^* e^{\bv\cdot\vecx^*})$. This can be used to define an objective function $E$ that will be used to determine the recruitment probability for the candidate
\begin{equation}
E(\vecx^*|D_t) = h(D_t) - H(\vecx^*|D_t).
\label{eq:objective}
\end{equation}

%°°°°°°°°°°°°°°°°°°°°°°°°°°°°°°°°°°°°°°%
\subsection{Mathematical approximations}
\label{sec:var}
%°°°°°°°°°°°°°°°°°°°°°°°°°°°°°°°°°°°°°°%

The expectation (\ref{eq:entropy}) is analytically intractable. Consequently, we develop a variational approximation of the the posterior $q(\lambda,\bv) \approx p(\lambda,\bv|D_t)$ with $q(\lambda,\bv) = q(\lambda)q(\bv)$. The purpose of a variational approximation is to approximate the posterior with a form that is more amenable to analytical integration \citep[Chapter 10]{bishop2006pattern}. For the variational distributions $q$ we choose a log-Normal distribution, $\log \lambda \sim \mathcal{N}(\mu_1,\sigma_1^2)$, and a multivariate Normal distribution for the regression coefficients, $\bv \sim \mathcal{N}(\mv_0,\Sigma_0)$ with $\Sigma_0=\text{diag}(\sigma^2_{01},\ldots,\sigma^2_{0d})$. To achieve a `good' approximation we minimise the Kullback-Leibler divergence between the distributions $q$ and $p$ with respect to the variational parameters $(\mu_1,\sigma_1^2,\mv_0,\sigma_{01}^2,\ldots,\sigma_{0d}^2)$:
\begin{align}
\text{KL}(q||p) &= \left<\log\left[\frac{q(\lambda)q(\bv)}{p(\lambda,\bv|D_t)}\right]\right>_{q(\lambda)q(\bv)}\nonumber\\
& = \left<\log q(\lambda)\right>_{q(\lambda)} + \left<\log q(\bv)\right>_{q(\bv)} - \left<\log p(\lambda,\bv|D_t)\right>_{q(\lambda)q(\bv)}.
\label{eq:kl}
\end{align}
This is convenient since the first two terms give the entropy of the variational distribution which is required in (\ref{eq:energy}). Equation (\ref{eq:kl}) is explicitly calculated in Appendix \ref{app:details}.

In addition, the expectations (\ref{eq:energy}, \ref{eq:pred}) are analytically intractable. We make two further approximations:
\begin{enumerate}
\item
$p(t^*|\vecx^*,\lambda,\bv) = \delta(t^*-\hat{t})$ where $\hat{t} = \left<t^*\right>_{p(t^*|\vecx^*,\lambda,\bv)} = (\lambda e^{\bv\cdot\vecx^*})^{-1}$.
\item $p(\lambda,\bv|D_t) = \delta(\hat{\lambda} - \lambda)\delta(\hat{\bv} - \bv)$ where $(\hat{\lambda},\hat{\bv}) = \text{argmax}_{(\lambda,\bv)} p(\lambda,\bv|D_t)$.
\end{enumerate}
The Dirac delta function $\delta(x)$ is loosely defined by $\delta(0)=\infty$ and is zero elsewhere. These approximations allow evaluation of the integrals (\ref{eq:energy}, \ref{eq:pred}) and, additionally, it is computationally faster to obtain $(\hat{\lambda},\hat{\bv})$ rather than numerically integrating (\ref{eq:energy}, \ref{eq:pred}). Combining the above approximations we can write $\hat{t} = (\hat{\lambda}e^{\hat{\bv}\cdot\vecx^*})^{-1}$ and obtain
\begin{align}
\hat{H}(\vecx^*|D_t) &=  \hat{h}(D_t\cup\{\vecx^*,\hat{t}\})\label{eq:approxE}\\
\hat{h}(D_t) &= -\left<\log q(\lambda)\right>_{q(\lambda)} - \left<\log q(\bv)\right>_{q(\bv)}\label{eq:approx}.
\end{align}
These can be substituted into (\ref{eq:objective}) to obtain an approximated objective function $\hat{E}(\vecx^*|D_t)$. Evaluation of these expressions require numerical optimisation of (\ref{eq:posterior}) and (\ref{eq:kl}) in order to evaluate, but this is computationally feasible. Note that estimates of $\lambda$ and $\bv$ could be unstable at the early stages of the trial when few patients have been recruited. In this case, one could implement a `burn in' phase where selective recruitment only begins after a certain number of patients have been recruited.

%°°°°°°°°°°°°°°°°°°°°°°°°°°°°°°°°°°°°°°%
\subsection{Obtaining a recruitment and allocation probability}
\label{sec:palloc}
%°°°°°°°°°°°°°°°°°°°°°°°°°°°°°°°°°°°°°°%

Once a candidate patient presents with covariates $\vecx^*$ we would like to define a recruitment probability $\rho(\vecx^*|D_t)$. In general, we can write $\vecx^* = [\vecy^*,\vecz]$ where $\vecy^*$ are clinical covariates or biomarkers and $\vecz$ indicates the allocated treatment arm. Suppose there are $K$ arms in total and $\vecz \in \{\vecz_1,\ldots,\vecz_K\}$ where $\vecz_{k}$ indicates allocation to arm $k$. The first step is to define the allocation probability to treatment arm $k$ as
\begin{equation}
p(k|\vecx^*,D_t) = \frac{\hat{E}(\vecy^*,\vecz_{k}|D_t)}{\sum_{j=1}^K  \hat{E}(\vecy^*,\vecz_{j}|D_t)}\quad\text{for $k=1,\ldots,K$.}
\label{eq:biased}
\end{equation}
A treatment arm is chosen at random according to this distribution and is denoted by $\vecz^*$. Secondly, we define the \emph{ideal} candidate as $\vecy_{I} = \text{argmax}_{\vecy} \hat{E}(\vecy,\vecz^*|D_t)$. The ideal candidate would give us the greatest reduction in expected entropy. A recruitment probability is given by
\begin{equation}
\rho(\vecx^*|D_t) = f_0\left(\frac{\hat{E}(\vecy^*,\vecz^*|D_t)}{\hat{E}(\vecy_I,\vecz^*|D_t)}\right)
\label{eq:rho}
\end{equation}
where $f_0$ is some function that remains to be specified. Since the argument of $f_0$ must lie in the interval $[0,1]$ we can choose $f_0$ to be the identity function in which case the closer the candidate is to the ideal patient the higher the probability of recruitment. Alternatively, we can choose $f_0(s) = \theta(s - p_0)$ for a specified threshold $p_0$. The step function $\theta(s) = 0$ if $s\leq0$ and $\theta(s)=1$ otherwise. This results in deterministic recruitment. A more general option is $f_0(s) = (1+\text{tanh}(s/\beta_0-p_0))/2$ which is equivalent to deterministic recruitment when $\beta_0\to0$. This allows the practitioner to implement a desired level of stringency in the recruitment process.

%ÇÇÇÇÇÇÇÇÇÇÇÇÇÇÇÇÇÇÇÇÇÇÇÇÇÇÇÇÇÇÇÇÇÇÇÇÇÇÇÇÇÇÇÇÇÇÇÇÇÇÇÇÇÇÇÇÇÇÇÇÇ%
%
%

%
%
%ÇÇÇÇÇÇÇÇÇÇÇÇÇÇÇÇÇÇÇÇÇÇÇÇÇÇÇÇÇÇÇÇÇÇÇÇÇÇÇÇÇÇÇÇÇÇÇÇÇÇÇÇÇÇÇÇÇÇÇÇÇ%

%°°°°°°°°°°°°°°°°°°°°°°°°°°°°°°°°°°°°°°%
\section{The German Breast Cancer Dataset}
\label{sec:results:gbcs}
%°°°°°°°°°°°°°°°°°°°°°°°°°°°°°°°°°°°°°°%

We applied our method to data obtained from the German Breast Cancer Study (GBCS) described in \citet[Section 1.3] {Hosmer2008}. Our goal is to infer the parameters for a single covariate in order to assess how our adaptive protocol performs. The data consist of time-to-event outcomes for 686 patients recruited between July 1984 and December 1989. There are eight covariates in total. We decided to use tumour size (mm) for a univariate analysis because a good spread (1st quartile $= 20$ mm, median = $25$ mm, 3rd quartile $=35$ mm) would make it suitable for filtering patients according to the covariate. Importantly, the dataset also contains the date at which each patient is diagnosed with primary node positive breast cancer so we can easily calculate the waiting-time between patients. This allows us to effectively `re-run' the trial. The primary event was recurrence.

\begin{figure}[t!]
\centering
\includegraphics[scale=0.65]{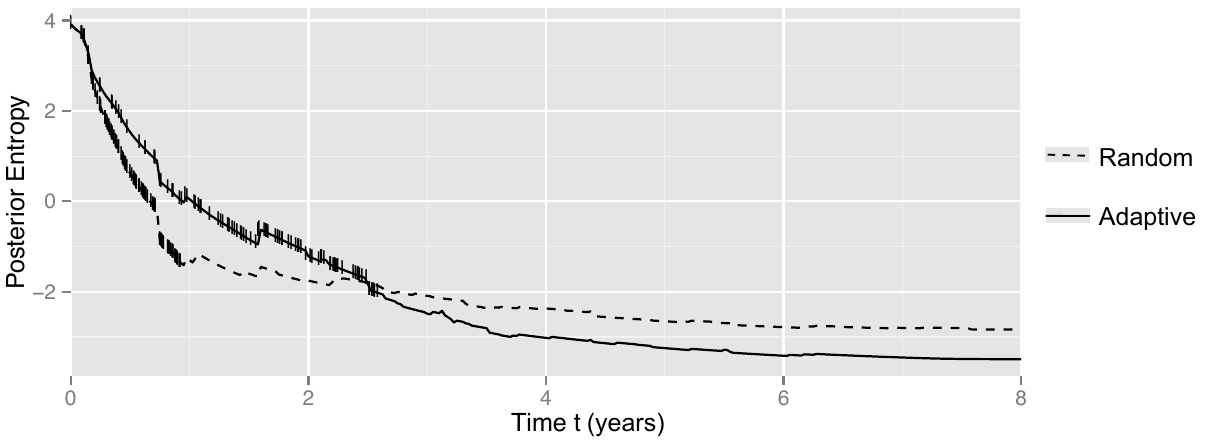}
\caption{Plot of the posterior entropy (\ref{eq:approx}) for the RCT and ACT as a function of time. The vertical ticks indicate times at which a patient was recruited. The sharp drop at $\approx0.75$ years corresponds to the first primary event occurring.}
\label{fig:entropy}
\end{figure}

To assess the information-adaptive design we decided to recruit a total of $N_{T}=100$ patients. We used deterministic recruitment with a cutoff of $p_0 = 0.5$. The trial was terminated after 10 years. We compared this to a randomised clinical trial (RCT) in which the first 100 patients are recruited. The same proportional hazards model as Section \ref{sec:ph} was used to analyse the RCT. The covariate values were median-centred and rescaled by 25 mm. The population density was assumed constant. We impose a uniform prior between $\pm1$ for the ideal covariate $x_{I}$. Hyperparameters were set to $(\kappa_0,\chi_0,\alpha_0^2)=(3,1,4)$.

%\begin{table}
%\caption{\label{tab:gbcs}Inferred parameters and entropies of the original GBCS dataset, the adaptive clinical trial (ACT), and  the randomised clinical trial (RCT). In brackets are 95 percent confidence intervals.}
%\centering
%\fbox{%
%%\begin{tabular}{*{4}{c}}
%\begin{tabular}{|c|c|c|c|}
%& $\lambda$ & $\beta$ & entropy\\
%\hline
%Full dataset ($N_{total}=686$) & 0.13 & 0.36 (0.19,0.52), p-value $=6.1\times10^{-6}$ & -4.54\\
%ACT ($N_{total} = 100$, $N_{reject}=278$) & 0.11 & 0.44 (0.21,0.66), p-value $=4.2\times10^{-5}$ & -3.49\\
%RCT ($N_{total} = 100$) & 0.14 & 0.11 (-0.27,0.48), p-value $=0.29$ & -2.83\\
%\hline
%\end{tabular}}
%\end{table}

\begin{table}
\centering
\begin{tabular}{|c|c|c|c|c|c|c|}
\hline
& $N_{total}$ & $N_{reject}$ & $t_R$ & $\lambda$ & $\beta$ & entropy\\
\hline
Full & 686 & 0 & 67 & 0.13 & 0.36 (0.19,0.52), $p=6.1\times10^{-6}$ & -4.54\\
ACT & 100 & 278 & 31 & 0.11 & 0.44 (0.21,0.66), $p=4.2\times10^{-5}$ & -3.49\\
RCT & 100 & 0 & 11 & 0.14 & 0.11 (-0.27,0.48), $p=0.29$ & -2.83\\
\hline
\end{tabular}
\caption{Inferred parameters and entropies of the full GBCS dataset (Full), the adaptive clinical trial (ACT), and  the randomised clinical trial (RCT). In brackets are 95 percent confidence intervals and $p$ is corresponding the p-value. $N_{total}$ is the total number of recruits, $N_{reject}$ is the number of rejected candidates, and $t_R$ is the recruitment time in months.}
\label{tab:gbcs}
\end{table}

It took approximately 1 year to recruit 100 patients onto the RCT. The adaptive clinical trial (ACT) took approximately 2.5 years, during which a total of 278 patients were rejected. In Figure \ref{fig:entropy} the posterior entropies for both the ACT and RCT are plotted. Initially the entropies are largely determined by the priors over $\lambda$ and $\beta$ but quickly drop as patients are recruited, although not monotonically. In the first 2.5 years of the trial the RCT has a lower entropy which is presumably due to the fact that more patients have been recruited compared to the ACT. Towards the end of the trial the ACT has a lower entropy due to a more informative cohort. Both entropies continue to decrease after recruitment has finished as more events are observed.

Table \ref{tab:gbcs} shows the inferred model parameters (evaluated after 10 years) from the original dataset, the ACT, and the RCT. The ACT results in a significant non-zero value for $\beta$ that is close to the value obtained using the full dataset (with $N=686$). The RCT fails to infer any significant value.

\begin{figure}[tb!]
\centering
\includegraphics[scale=0.5]{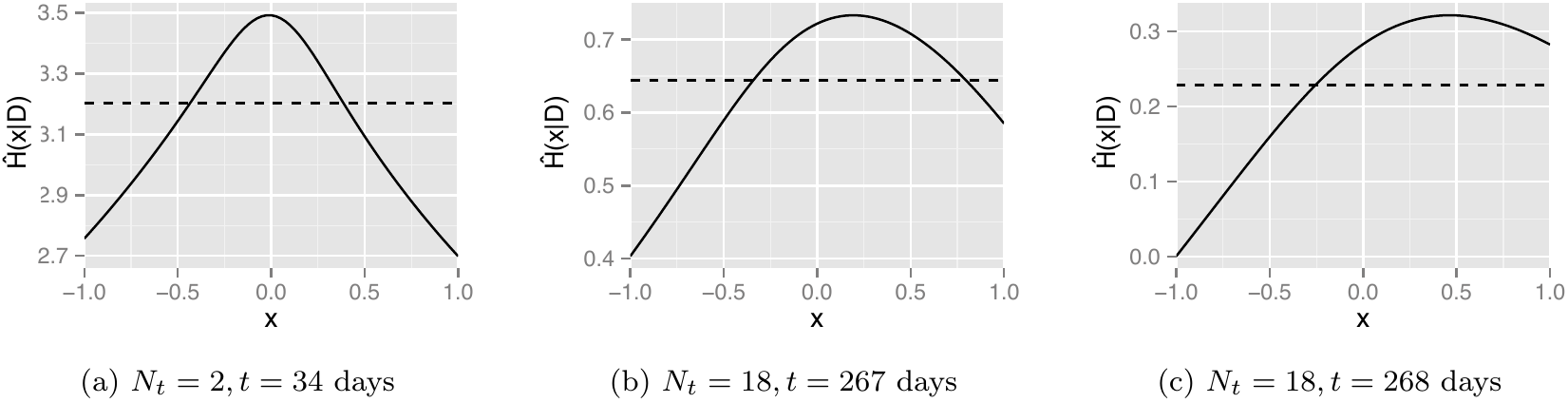}
\caption{The expected entropy (\ref{eq:approxE}) as a function of $x$ at various times during the ACT.}
% Generated from ~/ UCL/Adaptive trials/German brest cancer dataset/R/N_100_p_50
\label{fig:energy}
\end{figure}

In order to gain some intuition for how the recruitment probabilities are determined we have plotted the expected entropy as a function of the covariate $x$ at various time points in Figure \ref{fig:energy}. We note that the function tends to have one maximum and two minima at $x=\pm1$. This general shape is due to the nature of the proportional hazards model since extreme values of $x$ will diminish the space of plausible parameter values more so than values close to zero, and consequently are more informative. The dashed line is the entropy below which a candidate will be recruited. In (a) the trial has started at $t=0$ with two patients. There is a strong preference for individuals towards $\pm 1$. The next candidate (at $t=34$ days) had $x^*=-0.52$ and so was recruited. In (b), some patients with covariate values $>1$ have been recruited and this encourages recruitment of negative covariate values. At $t=267$ days no primary events have occurred. In (c), after $t=268$ days the first primary event occurs for a patient with a positive covariate value. This additional piece of information further increases the benefit of recruiting negative covariate values over positive ones. Note that the vertical scale changes. This illustrates that the recruitment probability changes dynamically, and depends on the observed events and covariate values of the existing cohort. We conclude that in general we gain more information from covariate values that have been under-sampled or values where few primary events have occurred.

\begin{figure}[t!]
\centering
\includegraphics[scale=0.6]{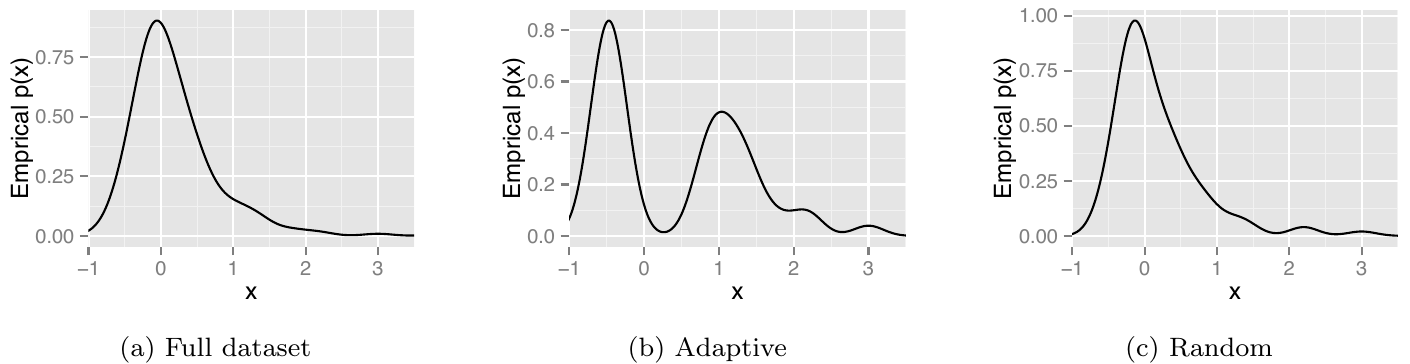}
\caption{Kernel smoothed empirical covariate densities (Gaussian kernel, bandwidth $=0.2$) for (a) the full GBCS dataset, (b) the ACT, and (c) the RCT.} 
% Generated from ~/ UCL/Adaptive trials/German brest cancer dataset/R/N_100_p_50
\label{fig:density}
\end{figure}

Individuals with covariates values far from zero will have the greatest reduction in expected entropy. This is because these terms will dominate the data likelihood in a proportional hazards model. Consequently, the covariate distribution in the ACT can differ considerably from the population distribution. Figure \ref{fig:density} shows the empirical covariate distributions for the original dataset and both trials. Due to the shape of the expected entropy function (see Figure \ref{fig:energy}) patients towards $\pm 1$ were more likely to be recruited in the ACT. Consequently, almost no patients with $x\approx 0$ were recruited. The RCT density resembles the density of the full dataset.

%°°°°°°°°°°°°°°°°°°°°°°°°°°°°°°°°°°°°°°%
\section{Numerical simulation studies}
\label{sec:results:num}
%°°°°°°°°°°°°°°°°°°°°°°°°°°°°°°°°°°°°°°%

Here we consider a scenario where the covariates consist of a two-dimensional biomarker $\vecy_i =(y_{i1},y_{i2})$ and patients are given one of three treatments denoted by $\vecz_i = (z_{i1},z_{i2},z_{i3})$. A patient given treatment one would have $\vecz_i = (1,0,0)$, treatment two would have $\vecz_i=(0,1,0)$, and so forth. We are interested in whether there is any interaction between the biomarker and treatments, i.e. is the biomarker predictive. A proportional hazards model with interaction terms is assumed:
\begin{equation}
h(t|\vecy_i,\vecz_i,\lambda,\bv) = \lambda e^{\beta_1y_{i1}z_{i1} + \beta_2y_{i1}z_{i2} + \beta_3y_{i1}z_{i3} + \beta_4y_{i2}z_{i1} + \beta_5y_{i2}z_{i2} + \beta_6y_{i2}z_{i3}}.
\label{eq:interaction}
\end{equation}
This gives a total of six regression coefficients and the baseline hazard $\lambda$ to be inferred. In all simulations we compared an adaptive trial to a randomised one.

To simulate survival data we generate a random vector $\vecy=(y_1,y_2)$ where $y_i\sim \text{uniform}(-1,+1)$ or $y_i\sim \mathcal{N}(0,0.5)$ for $i=1,2$. A treatment arm $\vecz$ is chosen (either randomly or according to (\ref{eq:biased})). A random number $w\sim \text{uniform}(0,1)$ is generated, and an event time is given by the inverse of the cumulative distribution $t = - e^{-\bv\cdot\vecx}\log(1-w)/\lambda$ where $\vecx\in\mathbb{R}^6$ contains the same product terms between $\vecy$ and $\vecz$ as (\ref{eq:interaction}). Patients are censored at random with probability $p_{c}\in[0,1]$. If an individual is censored then the time-to-censoring is drawn from a uniform density between 0 and $t$. The first patient to be generated is recruited onto both the ACT and RCT. The waiting time until the next patient is drawn from an exponential density with rate parameter $\xi$. Hyperparameters were set to $(\kappa_0,\chi_0,\alpha_0^2)=(3,1,4)$.

\begin{table}[b]
\centering
\begin{tabular}{|c|c|c|c|c|c|c|c|}
\hline
 & $\beta_1$ & $\beta_2$ & $\beta_3$& $\beta_4$& $\beta_5$& $\beta_6$ & $\lambda$\\
\hline
ACT (Uniform) & 0.348 & 0.374 & 0.361 & 0.384 & 0.418 & 0.352 & 0.00080\\
RCT (Uniform) & 0.364 & 0.347 & 0.401 & 0.389 & 0.396 & 0.384 & 0.00084\\
\hline
ACT (Gaussian) & 0.499 &0.5120 &0.487& 0.438& 0.445& 0.430& 0.00085\\
RCT (Gaussian) & 0.470 &0.494& 0.504& 0.471& 0.518& 0.435&0.00084\\
\hline
\end{tabular}
\caption{Mean square error between inferred and `true' model parameters over 500 simulations. Comparison between both random and adaptive trials without selective recruitment and uniform and Gaussian distributed covariates.}
\label{tab:ap1}
\end{table}

%°°°°°°°°°°°°°°°°°°°°°°°°°°°°°°°°°°°°°°%
\subsection{Adaptive allocation without selective recruitment}
%°°°°°°°°°°°°°°°°°°°°°°°°°°°°°°°°°°°°°°%

In these simulations all patients were recruited. A total of $N=50$ patients were recruited onto both trials. The trial was terminated after $t=100$ arbitrary units of time. The rate parameter for waiting times was $\xi=6$, and $p_{c}=0.5$. Model parameters were set to $\bv=(0.8,-0.5,1.1,-0.7,0.6,0.1)$ and $\lambda=0.1$. In the ACT the expected entropy was used to determine which treatment arm each individual was allocated to as described in Section \ref{sec:palloc}. In the RCT patients were allocated to one of the three arms at random.

A total of 500 simulations were run. We computed the mean square error between the inferred model parameters and the `true' values used to generate the data. As shown in Table \ref{tab:ap1} we found essentially no difference between the randomised and adaptive trials for either uniformly or Gaussian distributed covariates. We found that the entropy at the end of the ACTs with uniform covariates was on average slightly lower than the RCTs (2.14 and 2.20 respectively), although the difference was statistically significant (p-value $0.017$ with a one-sided paired t-test). For Gaussian distributed covariates the difference in entropies was insignificant. We also performed a chi-squared test to see if the allocation proportions of patients across arms differed from a uniform distribution. Each simulated trial was tested and we found no p-values less than 0.05 for either uniform or Gaussian distributed covariates. Since the chi-squared test was repeated for each trial the p-values were corrected for multiple hypothesis testing by controlling the false discovery rate (using the method of \citet{benjamini1995controlling}) with the `p.adjust' R function.

%°°°°°°°°°°°°°°°°°°°°°°°°°°°°°°°°°°°°°°%
\subsection{Adaptive allocation and recruitment}
%°°°°°°°°°°°°°°°°°°°°°°°°°°°°°°°°°°°°°°%

In these simulations the same parameters as above were used but patients were recruited onto the ACT selectively with a threshold of $p_0=0.66$. Over 500 simulations we found that the mean square error between the inferred and `true' parameters was considerably lower in the ACTs than the RCTs as shown in Table \ref{tab:ap}. For uniformly distributed covariates 48.9\% of the inferred parameter values were significant (at 0.05) in the ACT compared to 39.2\% in the RCTs. Furthermore, the mean entropy at the end of the ACTs was $0.93$, compared to $2.23$ in the RCTs. On average 140.7 (standard deviation 42.9) individuals are rejected.

In the case of Gaussian distributed covariates the difference is more pronounced. 50.4\% of parameters were significant in the ACT compared to 35.0\% in the RCT. An average of 240.0 patients were rejected (standard deviation 61.9). Due to the Gaussian distribution there are more patients in the less informative region around zero. Therefore the number of rejections is higher and the benefit more substantial.

We also explored the effect of the threshold $p_0$ on the trial results. When $p_0=0.33$ we found that the MSE (averaged over the six beta values) was 0.287 in the ACT compared to 0.372 in the RCT with 44.0\% of inferred parameters reaching statistical significance in the ACT compared to 39.6\% in the RCT. An average of 22.0 patients were rejected (standard deviation 6.45). When the threshold was increased to $p_0=0.90$ the MSE was 0.358 versus 0.363, and the proportion of significant parameters was 41.7\% versus 39.3\%, in the RCT and ACT respectively. On average 237.3 (standard deviation 86.5) patients were rejected. This suggests that setting the threshold too high can be counterproductive.

\begin{table}
\centering
\begin{tabular}{|c|c|c|c|c|c|c|c|}
\hline
 & $\beta_1$ & $\beta_2$ & $\beta_3$& $\beta_4$& $\beta_5$& $\beta_6$ & $\lambda$\\
\hline
ACT (uniform)& 0.324 &0.279& 0.313 &0.342 &0.279& 0.306 &0.00079\\
RCT (uniform)& 0.401& 0.335 &0.408& 0.375 &0.367 &0.361 & 0.00081\\
\hline
ACT (Gaussian) & 0.266 & 0.289 & 0.278 & 0.217 & 0.253 &0.262 &0.00085\\
RCT (Gaussian) & 0.444 & 0.553 & 0.509 & 0.521 & 0.502 &0.478& 0.00082\\
\hline
\end{tabular}
\caption{Mean square error between inferred and `true' model parameters over 500 simulations. Comparison between random and adaptive trials with selective recruitment.}
\label{tab:ap}
\end{table}

% Gaussian p(x) with mean 0 sd 0.5
%> mse.rct
%[1] 0.4438016 0.5531201 0.5090444 0.5208340 0.5017330 0.4779483
%> mse.act
%[1] 0.2658130 0.2891530 0.2777569 0.2165892 0.2531839 0.2616743
%> mse.lambda.act
%[1] 0.0008571713
%> mse.lambda.rct
%[1] 0.0008288692

%ÇÇÇÇÇÇÇÇÇÇÇÇÇÇÇÇÇÇÇÇÇÇÇÇÇÇÇÇÇÇÇÇÇÇÇÇÇÇÇÇÇÇÇÇÇÇÇÇÇÇÇÇÇÇÇÇÇÇÇÇÇ%
%
%
\section{Discussion}
\label{sec:disc}
%
%
%ÇÇÇÇÇÇÇÇÇÇÇÇÇÇÇÇÇÇÇÇÇÇÇÇÇÇÇÇÇÇÇÇÇÇÇÇÇÇÇÇÇÇÇÇÇÇÇÇÇÇÇÇÇÇÇÇÇÇÇÇÇ%

The practicality of our proposed design will depend on various economic and ethical considerations as well as the characteristics of each particular trial and the study population. For instance, if a covariate is relatively inexpensive to measure when compared to the costs of recruitment (treatment provision, follow-up, administration) then it may be sensible to selectively recruit informative patients. A large pool of patients can be inexpensively screened and then resources concentrated on those which are likely to provide the most information. In this case a selective recruitment design could result in significant cost reductions since fewer recruits are required overall.

Clinical trials are not primarily intended to be therapeutic, but rather as a means to generate medical evidence. Recruited patients may be exposed to treatments that are ineffective (e.g. a placebo) or that are possibly even harmful. Our proposed design offers the possibility to conduct a trial using fewer patients than a traditional randomised design. This may be ethically attractive in some cases since ultimately fewer patients are offered treatment options with uncertain efficaciousness. 

In a selective recruitment design the decision to recruit and allocate a patient can also take into account the probability of a successful response to treatment (although this was outside the scope of this paper). Patients can be recruited and allocated in a manner that balances the statistical informativeness of a decision against the potential benefit or harm to that individual. The decision making process must balance individual and collective benefits. Maximising statistical information offers a collective benefit to all patients outside the trial (both current and future) who could benefit from the trial findings. Naturally this must be offset by what is best for the trial participants. What our proposed design offers the practitioner is a framework to balance individual versus collective ethical considerations.

Selective recruitment designs suffer from a number of drawbacks, one of which is longer recruitment times. If the patient accrual rate is low it may render the overall recruitment period unfeasible. Selective recruitment designs are therefore only appropriate in situations where patients accrue relatively quickly or where longer recruitment periods are an acceptable compromise.

One of the consequences of a proportional hazards model is that the most informative patients tend to have extreme values of covariates. As a result the distribution of recruited patients may differ from the population distribution which might make it difficult to generalise results from the trial to the general population. Thus, some generalisability is sacrificed in return for greater statistical power. If this was deemed undesirable one could introduce a sufficient level of random sampling in addition to preferential accrual of informative patients. Each candidate patient has a minimum probability of recruitment with informative patients having a higher probability. Thus, selective recruitment need not be an all or nothing process; it can be used to enrich the trial with informative patients to a desired degree.

Finally, in the case of model misspecification undesirable biases may be introduced into the dataset because the model choice influences the covariate distribution considerably. An additional limitation is that it is not yet clear how to estimate the sample size required for a certain level of statistical power --- a calculation that is typically used when planning new trials.

In summary, our novel information-adaptive selective recruitment clinical trial design will reject non-informative patients. Individuals who are more likely to clarify the values of our model parameters are more likely to be recruited. We have demonstrated with both experimental and simulated data the feasibility of our approach. Statistically significant inferences can be achieved using fewer patients with a selective recruitment design than a randomised trial, although we found that treatment arm allocation using an entropy based measure (without selective recruitment) did not offer any improvement over a randomised design. Such a design may offer a more economical or ethically attractive route to discover the relationship between biomarkers, treatments, and survival outcomes.

It will be interesting to extend this work beyond the proportional hazards assumption to more complex survival models. Incorporation of response-adaptive protocols offer another promising extension. Throughout this work we have assumed a uniform population density. In the case of a non-uniform density it may be desirable to incorporate this into the definition of an ideal candidate such that an ideal candidate is both informative and likely to be observed. This will require further investigation. Further extensions of the model could include alternative outcomes such as binary or continuous measurements.

%ÇÇÇÇÇÇÇÇÇÇÇÇÇÇÇÇÇÇÇÇÇÇÇÇÇÇÇÇÇÇÇÇÇÇÇÇÇÇÇÇÇÇÇÇÇÇÇÇÇÇÇÇÇÇÇÇÇÇÇÇÇ%
%
%
\appendix
\section{Derivation of the Kullback-Leibler divergence}
\label{app:details}
%
%
%ÇÇÇÇÇÇÇÇÇÇÇÇÇÇÇÇÇÇÇÇÇÇÇÇÇÇÇÇÇÇÇÇÇÇÇÇÇÇÇÇÇÇÇÇÇÇÇÇÇÇÇÇÇÇÇÇÇÇÇÇÇ%

The first two terms of the Kullback-Leibler divergence (\ref{eq:kl}) in Section \ref{sec:var} are simply minus the entropies of the variational distributions. These are $\left<\log q(\lambda)\right>_{q(\lambda)} = -(1/2+\log(2\pi\sigma_1^2)/2+\mu_1)$ and $\left<\log q(\bv)\right>_{q(\bv)}= -\sum_{\nu=1}^d\log(2\pi e \sigma_{0\nu}^2)/2$. The third term from (\ref{eq:kl}) is
\begin{align}
-N^1_t\left<\log\lambda\right>_{q(\lambda)} - \Pv_t\cdot\left<\bv\right>_{q(\bv)}&+\left<\lambda\right>_{q(\lambda)}\sum_{i=1}^{N_t}t_i\left<e^{\bv\cdot\vecx_i}\right>_{q(\bv)}\nonumber\\
& - \left<\log p(\lambda|\kappa_0,\chi_0)\right>_{q(\lambda)} - \left<\log p(\bv|\alpha_0^2)\right>_{q(\bv)}
\end{align}
where $N_t^1$ is the number of non-censored events up until time $t$ and $\Pv_t = \sum_{i:\Delta_i=1}\vecx_i$. It is straightforward to show $\left<\log\lambda\right>_{q(\lambda)} = \mu_1$, $\left<\lambda\right>_{q(\lambda)} = e^{\mu_1+\sigma^2_1/2}$ and $\left<\bv\right>_{q(\bv)} = \mv_0$. The following result is needed \citep[Appendix D]{Coolen2005}:
\begin{equation}
\int\text{d}\vecz\, \frac{e^{-\frac{1}{2}(\vecz-\mv)\cdot A^{-1}(\vecz-\mv) + \vecb\cdot\vecz}}{(2\pi)^{d/2}|A|^{1/2}} = e^{\mv\cdot\vecb + \frac{1}{2}\vecb\cdot A\vecb}
\label{eq:mgf}
\end{equation}
from which it follows $\left<e^{\bv\cdot\vecx_i}\right>_{q(\bv)} = e^{\mv_0\cdot\vecx_i + \frac{1}{2}\vecx_i\cdot\Sv_0\vecx_i}$. Note that (\ref{eq:mgf}) also defines the moment generating function for a multivariate normal distribution with mean $\mv$ and covariance matrix $A$. The terms relating to the priors are $\left<\log p(\bv|\alpha_0^2)\right>_{q(\bv)} 
=-\sum_{\nu}(\sigma^2_{0\nu} + [\mv_0]_{\nu}^2)/2\alpha_0^2$ and $\left<\log p(\lambda|\kappa_0,\chi_0)\right>_{q(\lambda)} = (\kappa_0-1)\left<\log\lambda\right>_{q(\lambda)} - \chi_0^{-1}\left<\lambda\right>_{q(\lambda)}$ where $[\mv_0]_{\nu}$ denotes the $\nu$th component of $\mv_0$.

\bibliographystyle{plainnat}
\bibliography{refs}

\end{document}